\documentclass[12pt]{article}

\usepackage[utf8]{inputenc}
\usepackage{fullpage, amsmath}
\usepackage[square,numbers,sort&compress]{natbib}

\usepackage{graphicx}
\usepackage{hyperref}
\usepackage{lineno}

\usepackage{authblk}

\title{The role of the mathematical sciences in supporting the COVID-19 response in Australia and New Zealand}
\date{}

\author[1,2]{James M. McCaw}
\author[3]{Michael J. Plank}
\affil[1]{School of Mathematics and Statistics, University of Melbourne, Melbourne Australia}
\affil[2]{Infectious Disease Dynamics Unit, Centre for Epidemiology and Biostatistics, Melbourne School of Population and Global Health, University of Melbourne, Melbourne Australia}
\affil[3]{School of Mathematics and Statistics, University of Canterbury, Christchurch, New Zealand}

\begin{document}
\maketitle

\clearpage

\begin{abstract}
Mathematical modelling has been used to support the response to the COVID-19 pandemic in countries around the world including Australia and New Zealand. Both these countries have followed similar pandemic response strategies, using a combination of strict border measures and community interventions to minimise infection rates until high vaccine coverage was achieved. This required a different set of modelling tools to those used in countries that experienced much higher levels of prevalence throughout the pandemic. 

In this article, we provide an overview of some of the mathematical modelling and data analytics work that has helped to inform the policy response to the pandemic in Australia and New Zealand. This is a reflection on our experiences working at the modelling-policy interface and the impact this has had on the pandemic response. We outline the various types of model outputs, from short-term forecasts to longer-term scenario models, that have been used in different contexts. We discuss issues relating to communication between mathematical modellers and stakeholders such as health officials and policymakers. We conclude with some future challenges and opportunities in this area. 

\end{abstract}

\clearpage

\section{Introduction}

The COVID-19 pandemic has posed major public health, policy and economic challenges to governments around the world. In navigating these challenges, governments have drawn on a range of scientific advice including from public health, infectious disease epidemiology, genomics, immunology, social science, and mathematical sciences. 

Australia and New Zealand have taken similar approaches to managing the pandemic, with both countries rapidly enacting strict border controls to minimise imported cases, and pursuing aggressive suppression through strict public health and social measures, until high vaccination coverage was achieved. Despite numerous border-related incursions of varying sizes \citep{douglas2021real,grout2021failures}, community transmission of SARS-CoV-2 was eliminated entirely from New Zealand and from the majority of states and territories of Australia for long periods of 2020 and 2021 \citep{baker2020successful,price2020early,golding2023modelling}. Both countries achieved high vaccination rates in late 2021, enabling them to adopt a revised COVID-19 strategy, progressively relaxing border restrictions and community interventions while keeping anticipated health impacts manageable. As a result, levels of pandemic mortality in both countries, whether measured through confirmed COVID-19 deaths or excess mortality, were among the lowest in the world. 

Mathematical modelling has played a significant role in both Australia and New Zealand in informing the policy and operational response to the pandemic. Because of the extended period of time under elimination or strict suppression strategies, this has required the development of a different set of modelling tools to those used in countries where there were significant levels of community transmission continuously since early 2020. For example, there has been a focus on modelling the effect of border controls \citep{shearer2022rapid,steyn2021managing} and an intensive test-trace-isolate-quarantine (TTIQ) system \citep{shearer2023estimating,james2021successful} as key interventions within an elimination or strict suppression strategy. There has also been a need for tools that provide situational awareness when prevalence is very low or zero. In these settings, traditional epidemiological data, for example on reported cases, is insufficient to provide a comprehensive view of the epidemic situation and public health priorities \citep{golding2023modelling}.

In this article, we will review some of the modelling tools used in Australia and New Zealand for policy advice and describe the impact these have had on the pandemic  response. We will compare and contrast the approaches taken in the two countries, identifying strengths and areas where increased collaboration could potentially be beneficial. We will explore how the characteristics and requirements of those tools changed with the establishment of widespread community transmission following the emergence of the Omicron variants of SARS-CoV-2 in late 2021 \citep{viana2022rapid}. We will also discuss how modelling groups have interacted with government officials and policymakers and identify some challenges and examples of good practice.

Our review and evaluations should not be considered systematic or exhaustive, but rather reflective of our own experiences as applied mathematicians working in leadership roles supporting our governments' responses to COVID-19. The work described is the outcome of the dedication and efforts of a large number of mathematicians and statisticians collaborating with epidemiologists and public health agencies. Much of the work has been carried out by early career researchers (ECRs), often employed on short-term contracts. Although not the main focus of this article, consideration of the perspectives of ECRs, and the suitability of academic reward systems and funding structures for supporting the delivery of urgent policy advice have been acknowledged \cite{kucharski2020therole,sherratt2023improving} and are important areas for further reflection and action.

\section{Modelling tools used in Australia and New Zealand} \label{sec:modelling_tools}

\begin{figure}
\centering
\includegraphics[trim={3cm 21cm 6cm 2cm},clip,width=\textwidth]{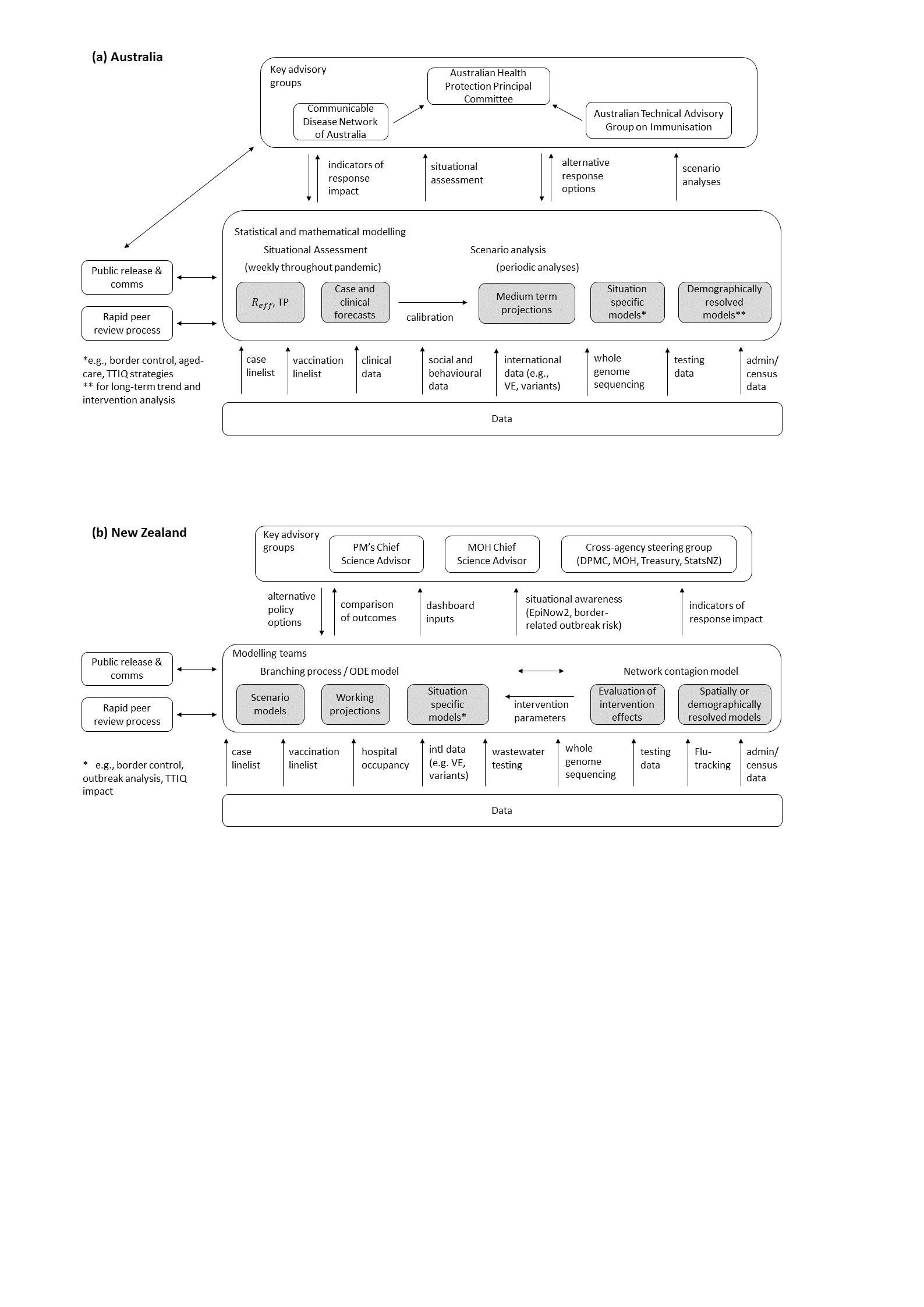}
\caption{Schematic diagrams showing key data requirements, workflows and stakeholder relationships for modelling groups in Australia and New Zealand. These are not official organisational charts, but rather a subjective representation of how some of the various modelling activities described in the text (central box) have used a range of data streams (bottom box) and interacted with government advisory groups (top box) at the data-modelling-policy interface. Abbreviations: DPMC -- Department of the Prime Minister and Cabinet; MOH -- Ministry of Health; $R_\mathrm{eff}$ -- effective reproduction number; ODE -- ordinary differential equation; TP -- transmission potential; TTIQ -- test-trace-isolate-quarantine; VE -- vaccine effectiveness. Panel (a) adapted from original produced by Dr Freya Shearer. }    
\label{fig:org_charts}
\end{figure}

Australia has used short-term statistical and mechanistic forecasting models to provide regular weekly situational assessment advice to government \citep{price2020early,golding2023modelling,moss2022forecast,IDDU2020techreport1,IDDU2020techreport2,IDDU2020techreport3,IDDU2020techreport4,IDDU2021techreport,IDDU2022techreport} based on real-time data on cases, behaviour and vaccination coverage. Outputs were agreed upon through formal processes of government, documented in Australia's National Surveillance Plan for COVID-19 \citep{AUGov2022surveillanceplan}, and a national consortium provided rapid internal peer-review of methods and outputs each week. These activities have been complemented by the periodic use of strategic scenario modelling for longer-term planning at key points in the pandemic. These include the onset of the pandemic in January 2020 \citep{moss2020coronavirus}, and when Australia began its transition to ``living with COVID'' in mid-to-late 2021 based on achieving high vaccination coverage \citep{conway2023covid,shearer2023estimating,ryan2022estimating,Doherty2021natplanwebsite} -- see Figure \ref{fig:org_charts}a.

New Zealand has primarily used mechanistic transmission models, fed with real-time data on vaccination rates, to provide advice to government, particularly at key points when policy changes were considered or implemented. Where important parameters were uncertain or context-dependent, an approximate Bayesian computation (ABC) approach was used to calibrate models to epidemiological data. Models have been periodically updated as needed, for example with the arrival of new variants, new data on measures such as vaccine effectiveness, or when epidemic dynamics began to clearly deviate from model projections. Reports describing model assumptions and results were initially informally reviewed by a multi-disciplinary rapid peer-review panel and publicly released as pre-prints concurrently with the policy decisions they were designed to support (see \url{www.covid19modelling.ac.nz/reports}). 

The rapid peer-review processes used in Australia and New Zealand provided some level of quality assurance for results being used for policy advice in real time. This was complemented with traditional scientific peer review and publication over longer timescales. This allowed a robustly peer-reviewed methodological model base to be built up over time, which could be rapidly applied to different contexts and policy questions as the need arose.

Both countries drew on, and in places extended, analyses and/or models published by researchers in the UK's Scientific Pandemic Influenza Group on Modelling (SPI-M-O) \citep{davies2020age,davies2021increased,davies2021estimated,moore2021vaccination,keeling2022comparison,barnard2022modelling,keeling2021short} and international data and analyses published by the UK Health Security Agency and others \citep{IDDU2021techreport,verity2020estimates,herrera2022age,andrews2022covid,nyberg2022comparative}. In this section, we give a description of some of the key model developments and examples of how they were used in each country.

\subsection{Australia} \label{sec:aus}

\subsubsection{Early scenario modelling to support initial planning and response}

In mid January 2020, Professor Jodie McVernon (Director of Epidemiology, Peter Doherty Institute) and Professor James McCaw (Head, Infectious Disease Dynamic Unit (IDDU), University of Melbourne) were engaged by the Australian Government Department of Health and Ageing to advise on the emerging threat from the virus now known as SARS-CoV-2. By the first week of February 2020, with few recorded cases globally and only preliminary information on the rate of transmission from early studies in Wuhan \citep{li2020early,kucharski2020early}, they had drawn on earlier work on pandemic influenza and SARS (that is, SARS-CoV-1) to develop a simple SEIR model to explore the potential impacts of COVID-19 in Australia. They used this to evaluate how measures such as case isolation, quarantine of contacts, and modest levels of social distancing may suppress transmission.

A key element of the early modelling work in Australia was a preliminary epidemiological investigation conducted in China \citep{du2020serial} which indicated that the serial interval distribution included probability mass at negative values. The serial interval is the time between symptom onsets in an infector-infectee pair and density at negative intervals is highly suggestive of pre-symptomatic transmission, now well understood to be a key feature of SARS-CoV-2 transmission.

\begin{figure}
    \centering
    \includegraphics[width=0.9\textwidth]{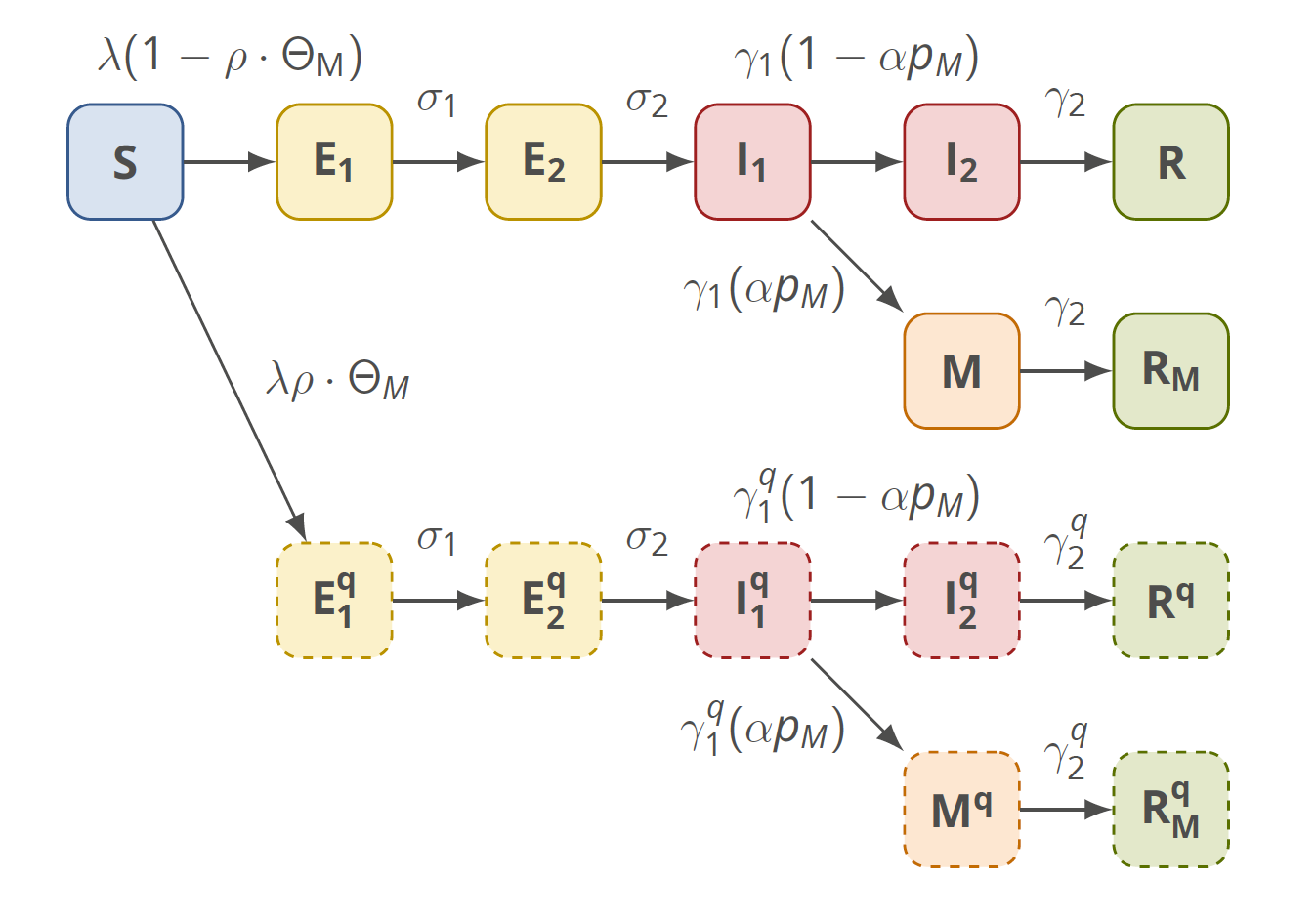}
    \caption{Schematic diagram of the deterministic transmission model used in Australia in early 2020. Compartments are: susceptible ($S$), exposed ($E_1$, $E_2$), infectious ($I_1$, $I_2$), `managed' ($M$) and recovered ($R$). A proportion $p_M$ of presenting cases (themselves a proportion $\alpha$ of all infections) are ascertained and isolated (compartment $M$). Quarantined persons are shown as compartments with superscript $q$ (dashed borders). Managed and quarantined compartments exert a lesser force of infection than non-managed, non-quarantined compartments.  Full mathematical details, including the expression for the force-of-infection $\lambda$, and process by which susceptible persons ($S$) are routed to the quarantine pathway (dashed borders), are provided in the Appendix of \citep{moss2020coronavirus}.}
    \label{fig:Moss-COVID2020}
\end{figure}

The model included two key elements beyond the foundational SEIR equations (Figure~\ref{fig:Moss-COVID2020}). Firstly, it included compartments for those who were contagious but not yet displaying symptoms, allowing the interplay between case-based isolation and pre-symptomatic transmission to be explored. And secondly, it modelled contacts, allowing for quarantine strategies to be evaluated. This rapid model-based risk assessment was completed by extending on earlier work \citep{mccaw2007prophylaxis} that introduced a method for modelling contacts in a simple ordinary differential equation (ODE) framework, which had laid the foundations for around 15 years of pandemic preparedness modelling in Australia.

The value of this early work \citep{moss2020coronavirus} was not in its quantitative findings. Rather, it demonstrated that case isolation and quarantine, particularly when combined with moderate social distancing, could achieve a valuable reduction in the rate of transmission and effectively, perhaps dramatically, ``flatten the curve''. The work provided senior health officials with a clear reason to advise government to act. Measures that were considered feasible and appropriate --- a modest reduction in contacts of around 25\%--33\% and enactment of quarantine and isolation policies --- were expected to have an substantive, potentially game-changing, impact on the epidemic.

As history records, action was taken and by late March 2020 --- supported by strict border closures (also informed by model-based risk assessment \citep{shearer2022rapid}) that slowed the rate of importation and strong (as opposed to the modest as modelled) social distancing measures enacted --- transmission in Australia was in decline, leading to predicted elimination of local transmission \citep{price2020early}.

\subsubsection{Real-time analysis, transmission potential and forecasts: COVID-19 situational assessment}

Elimination of SARS-CoV-2 from circulation presented a major methodological challenge for infectious disease dynamics. Best practice for monitoring transmission is to estimate the effective reproduction number $R_\mathrm{eff}(t)$. At the heart of any method to compute $R_\mathrm{eff}(t)$ is the renewal equation:
\begin{equation} \label{eq:renewal}
\text{Inc}(t) = R_\text{eff}(t) \int_0^{\infty} \text{Inc}(t-\tau) f(\tau) \; d\tau,
\end{equation}
where $\text{Inc}(t)$ is the incidence of infection at time $t$ (i.e., rate of new infectives) and $f(\tau)$ is the generation interval density function. As Equation~(\ref{eq:renewal}) implies, there is no need for all infections to be recorded as cases: recording of a constant fraction of infections is sufficient to estimate $R_\text{eff}(t)$.

Evaluation of $R_\text{eff}(t)$ --- a key metric for risk-assessment --- clearly requires established transmission, and the recording of incident cases in health databases. With local (but of course, temporary) elimination achieved, there was no way to make such a risk assessment. However, the ability for a respiratory pathogen, such as SARS-CoV-2, to spread through the population depends on a number of behavioural factors --- who we mix with and how we act during encounters. 

In April 2020, with temporary elimination achieved, Professor Nick Golding from Curtin University worked in collaboration with members of the IDDU at the University of Melbourne to develop a novel metric, the ``transmission potential'' (TP), to provide risk-assessment in the absence of a computable estimate of $R_\text{eff}(t)$. The semi-mechanistic approach \citep{golding2023modelling} draws on behavioural data streams (and case data where available) to provide an assessment of risk during periods of zero, low and high transmission. Conceptually, this method can be considered a generalisation of Bayesian methods for estimating $R_\text{eff}(t)$, where behavioural time-series data are used to compute a time-dependent prior for $R_\text{eff}(t)$ (the TP). In the absence of case data (because it is not available or there are no infections occurring in the community), we interpret that prior (the TP) as an estimate of the ability of the virus, if it were present, to spread in the population. During periods of virus circulation, the methodology provides an estimate of $R_\text{eff}(t)$ and the discrepancy between the prior (TP) and posterior ($R_\text{eff}$) quantifies the difference between the expected and realised transmission. This discrepancy is directly interpretatble through an epidemiological and public health lens \citep{golding2023modelling}.

The other major element of situational assessment is forecasting. In Australia, Dr Rob Moss (IDDU) and Professor James McCaw had worked in collaboration with Defence Science Technology Group (DSTG) and public health authorities for many years to establish a seasonal influenza forecasting capability. Times-series data from case-based surveillance systems were analysed with SIR-type transmission models using a Bayesian particle filter \citep{moss2016forecasting,moss2017retrospective}, with extensions to include evaluation of seasonality \citep{zarebski2017model}, healthcare-seeking behaviour \citep{moss2019accounting} and the role of forecasting in public health decision-making processes \citep{moss2018epidemic,moss2019anatomy}. Combined with US Defence-funded research in pandemic decision support \citep{shearer2020infectious,shearer2021development}, led by Dr Freya Shearer (IDDU), this provided the foundation from which Australia's national COVID-19 forecasting capability was rapidly developed. By mid 2020, the national situational assessment consortium had established a suite of models for producing weekly month-ahead ensemble forecasts of COVID-19 case incidence. A team at Monash University contributed a time-series analysis statistical model and a team at the University of Adelaide produced a stochastic branching process model. In Melbourne, the EpiFX platform used for seasonal influenza forecasting was adapted for COVID-19 \citep{moss2022forecast}. From 2022, DSTG contributed an additional compartmental model to the ensemble forecast. Over the course of the pandemic, numerous enhancements to the case forecasting models have been made including the role and impact of: vaccination and waning immunity on short- and medium-term transmission dynamics; variants of concern, that is new strains of SARS-CoV-2 with enhanced intrinsic transmissibility and/or immune-escape properties; anticipated or modelled changes in mixing patterns and behaviour; and changes in case ascertainment due to, for example, changes in testing practices.

Forecasting clinical loads, and in particular hospital ward and ICU occupancy levels, was also a major focus. In early 2020, Dr David Price and Dr Freya Shearer used estimates of admission rates and length-of-stay statistics to map from case-incidence forecasts to hospital ward and ICU occupancy \citep{price2020early}. A more sophisticated approach, accounting for age- and jurisdictional-specific factors, as well as variants, was developed over 2021 and 2022 in the IDDU, informed by a detailed epidemiological study on length-of-stay statistics in New South Wales \citep{tobin2023real-time}. The full details of those clinical forecasting methods are being prepared for scientific peer-review.

\subsubsection{Australia's national re-opening plan 2021}

While not detailed here, major undertakings in scenario analyses were conducted in mid- to late-2021 to support Australia's ``national re-opening plan'', a major shift in Australia's strategic response to COVID-19. That high-impact work, co-led by McVernon and McCaw and delivered by a large national consortium \citep{Doherty2021natplanwebsite}, is not reviewed here. Details are available in a number of technical reports (see aforementioned website) and publications \citep{conway2023covid,shearer2023estimating,ryan2022estimating}.

\subsection{New Zealand} \label{sec:nz}

\begin{figure}
    \centering
    \includegraphics[width=\textwidth]{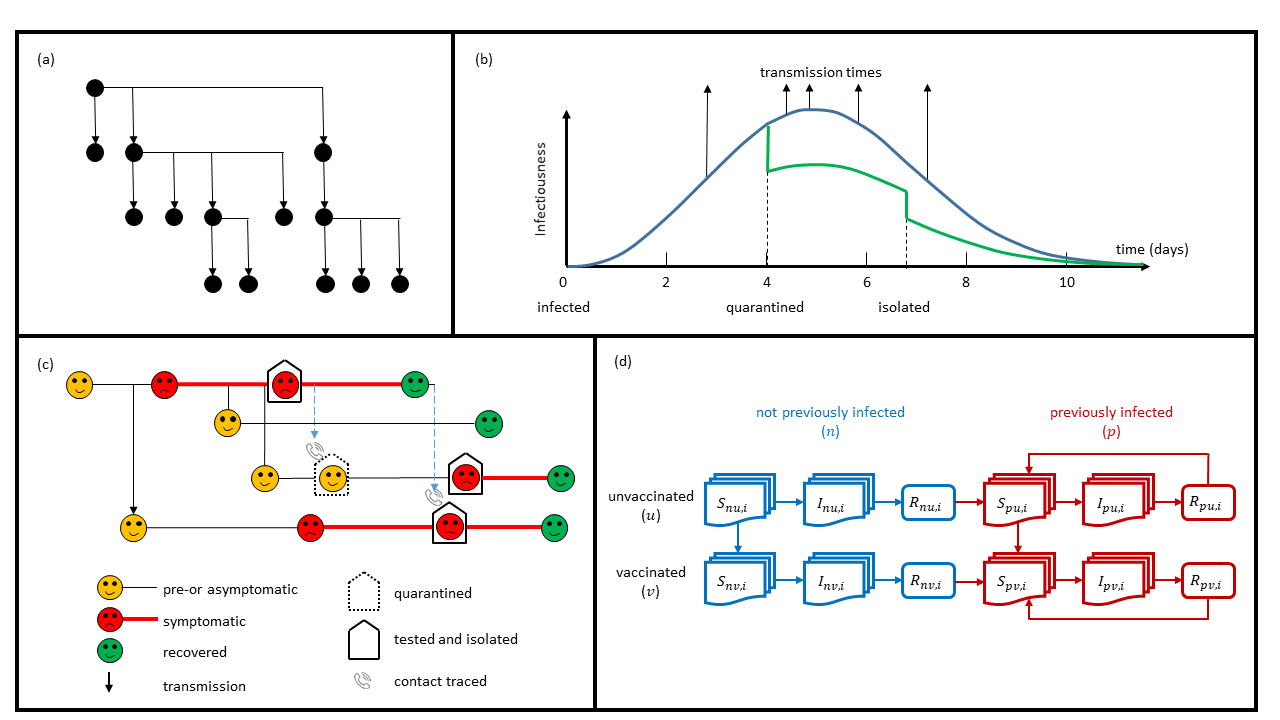}
    \caption{Schematic diagram of the stochastic nonlinear branching process model used in New Zealand. (a) A branching process model generates an explicit transmission tree and allows for a right-skewed offspring distribution, where a high proportion of individuals infect few others and transmission is dominated by a minority of superspreaders. This was incorporated into the model by assigning each infected individual a random transmission rate multiplier $Y_i$ drawn from a gamma distribution. (b) Individual transmission rate over time is governed by a generation time distribution $w(t)$ (blue); if an individual is quarantined or isolated their transmission rate is reduced (green). (c) Simplified test-trace-isolate-quarantine model: symptomatic individuals have a prescribed probability of testing, with some time delay from symptom onset to test; confirmed cases are isolated and their contacts are traced and quarantined with a prescribed probability, with some time delay from confirmation of the index case to quarantine of the contact. (d) Diagram of main model compartments: the susceptible population $S_i$ in age group $i$ is split according to vaccination status and prior infection (susceptible compartments are further divided to account for number of vaccine doses and waning immunity --- not shown here). Each susceptible compartment is associated with different levels of immunity against infection, hospitalisation and death. Colour available online.}
    \label{fig:BPM_schematic}
\end{figure}

\subsubsection{Early scenario modelling to support initial response and elimination strategy}

In March 2020, a group of researchers at Te P\=unaha Matatini (New Zealand's Centre of Research Excellence in Complex Systems), led by Professor Shaun Hendy, used simple SEIR models to estimate the potential scale of the health impact of COVID-19 in New Zealand under different interventions \citep{james2020suppression}. Interventions were modelled in a very simplistic way and their effectiveness was highly uncertain at the time. Nevertheless, the models were sufficient to show that a rapid and highly effective response would be needed to avoid quickly overwhelming the healthcare system and thousands of deaths. These results were communicated via the Prime Minister's Chief Science Advisor and Ministry of Health Chief Science Advisor and helped to inform decisions to close the border and, on 23 March 2020, to instigate a strict nationwide ``Alert Level 4'' lockdown \citep{hendy2022integrating}. 

Two main modelling tools were subsequently developed by groups at COVID-19 Modelling Aotearoa (formerly Te P\=unaha Matatini): a network contagion model, led by Dr Dion O'Neale and Dr Emily Harvey, and a nonlinear branching process model (and later a compartment-based ODE model), led by Professor Michael Plank and Professor Alex James. Both models were stochastic: New Zealand's very low case numbers in 2020-21 meant stochastic effects were important and deterministic models were less useful. 

The network contagion model is an agent-based model that simulates transmission on a synthetic contact network representing New Zealand's population. The network was constructed from individual-level employment and education micro-data in New Zealand's Integrated Data Infrastructure \citep{statsnz_idi}, with separate network layers for interactions occurring in different settings (home, work, school and community) \citep{harvey2020network}. 

Most of this section focuses primarily on the branching process model (the workstream led by MP). The model is similar in structure to a stochastic SEIR model, but allows for non-Markovian transitions and individual variability among infectious individuals (Figure \ref{fig:BPM_schematic}). The most basic form of the model can be expressed as the mean number of individuals infected by individual $i$ at time $t$:
\begin{equation} \label{bpm_equation}
\Lambda_i(t) = R_0 Y_i F_i(t) C(t) \hat{f}\left(t-t_{\mathrm{inf},i} \right)  s(t), \end{equation}
where $R_0$ is the basic reproduction number, $Y_i$ is a random variable representing individual heterogeneity in transmission \citep{lloyd2005superspreading}, $F_i(t)$ represents the effect of isolation or quarantine on individual $i$ at time $t$, $C(t)$ represents the effect of social distancing measures at time $t$, $\hat{f}$ is the probability mass function for the generation interval, and $s(t)$ is the proportion of the population that is susceptible at time $t$.

Having a combination of modelling approaches proved useful. The branching process model was less computationally intensive, required fewer parameter assumptions, and could be readily adapted to a variety of policy questions, for example how testing policies for international arrivals and hotel quarantine workers affected the risk of border-related outbreaks \citep{steyn2021managing,plank2021vaccination}. However, it only offered a high-level picture. The network model provided a more detailed view, for example the ability to provide spatial outputs, to stratify by ethnicity, employment type or deprivation index, and to explicitly model household and non-household transmission \citep{harvey2020network}. Some policy questions were more suitable for one model than the other and often a summary of complementary outputs from both models was communicated to policymakers. In some instances, the network model was used to simulate the relative effect of a specific intervention \citep{harvey2021contagion} and the results were used to estimate parameters for scenario analysis in the branching process model \citep{vattiato2023modelling} -- see Figure \ref{fig:org_charts}b.

By April 2020, New Zealand had formally adopted an elimination strategy \citep{baker2020new}, aiming to stamp out community transmission and use border controls to prevent re-introduction. Between April and June 2020, simulations of the branching process model were run with seed infections representing reported travel-related cases. These results were used to investigate the potential consequences of relaxing restrictions, informed by international estimates of the effects of various non-pharmaceutical interventions \citep{flaxman2020estimating,hsiang2020effect}. The model was also used to estimate the probability of elimination, conditional on fitting observed data on daily reported cases \citep{hendy2021mathematical}.

\subsubsection{Modelling border controls and resurgent outbreaks}

On 8 June 2020, the New Zealand government declared that community transmission of SARS-CoV-2 had been eliminated and removed all social distancing requirements \citep{nzgovt2020AL1}. On several occasions between June 2020 and August 2021, a case of COVID-19 was detected with no known link to the border \citep{douglas2021real}. In these situations, the branching process model was used to estimate the potential number of undetected community cases. New Zealand lacked a computable estimate comparable to Australia's transmission potential, a clear gap in situational awareness capability. Instead, simulations were run across a range of assumptions for $R_\mathrm{eff}(t)$, as well as the probability of symptomatic community testing and the number of transmission chains from the index to the detected case \citep{steyn2021managing}. 

When a new case was detected in Auckland on 11 August 2020, based on the fact that there was no known link to the border, the model estimated there were 10--50 people already infected. This prompted an immediate decision to move to ``Alert Level 3'' in Auckland, which, coupled with an intensive test-trace-isolate-quarantine (TTIQ) operation, ultimately eliminated the outbreak after 179 confirmed cases \citep{hendy2022integrating}. 
In November 2020, another unlinked case was detected. However, whole genome sequencing subsequently suggested a link to another case known to have been exposed at a hotel quarantine facility. In this case, the model estimated there were likely to be 4--20 other infections. This result, combined with new estimates of the effect of TTIQ on transmission \citep{james2021successful}, helped avoid another costly lockdown \citep{nzgovt2020covid,hendy2022integrating}. The cluster was subsequently eliminated via TTIQ after 6 confirmed cases \citep{douglas2021real}.

\subsubsection{Modelling the vaccine rollout and shift away from elimination}

Over time, the branching process model was developed to include additional variables and processes such as age structure, vaccination status, waning immunity and the effect of different SARS-CoV-2 variants (Figure \ref{fig:BPM_schematic}d). This work drew on models developed by the Infectious Disease Epidemiology group at the University of Warwick \citep{moore2021vaccination,dyson2021possible,keeling2021waning} and made similar assumptions to the subsequently published Doherty National Plan Modelling in Australia \citep{conway2023covid}. Age structure was modelled via a contact matrix, derived from pre-pandemic social survey data from European countries \citep{mossong2008social,prem2017projecting} and projected onto New Zealand's population. 

The age-structured model was used to estimate the reduction in $R_\mathrm{eff}(t)$ and potential health impacts at different stages of the vaccine rollout \citep{steyn2022covid}. By early 2021, a cross-agency modelling steering group had been established to liaise between modellers and policy makers, although the Chief Science Advisors were still playing an important role in communicating results to decision makers (see Figure \ref{fig:org_charts}b). The results of the model showed that, with the B.1.617.2 (Delta) variant becoming dominant globally, it would not be possible to reach herd immunity by vaccination alone due to Delta's higher transmissibility. This implied that additional public health measures would likely be needed to control the epidemic and manage the transition from elimination to living with the virus. These results also informed decisions around the progressive relaxation of border restrictions \citep{steyn2022effect}.

Previous studies showed that M\=aori and Pacific People were likely to be at higher risk of severe illness and death as a result of COVID-19 after controlling for age \citep{steyn2020estimated,steyn2021maori}. M\=aori have a younger age demographic and, combined with the higher risk, this suggested that the age-based vaccine rollout strategy should be adjusted so that M\=aori would become eligible to receive the vaccine at a younger age. However, the Crown chose not to implement this recommendation, a decision that was ultimately found to be in breach of Te Tiriti o Waitangi \citep{waitangi2021haumaru}.

In August 2021, New Zealand experienced an outbreak of the Delta variant, most likely linked to a traveller from New South Wales  \citep{jelley2022genomic}. At the time, around 32\% of the population had received at least one dose of the Pfizer/BioNTech vaccine and 19\% had received two doses. Our model estimated that vaccination was providing approximately a 15\% reduction in $R_\mathrm{eff}(t)$ and that, without a stringent response, healthcare systems would likely be overwhelmed within weeks \citep{plank2022using}. Outbreak size and risk of inter-regional spread were also estimated with the network model \citep{gilmour2021modelling,gilmour2021inter}. For the first time since March 2020, the government imposed a nationwide Alert Level 4 lockdown. Although this ultimately failed to eliminate the outbreak, it kept it geographically localised to Auckland and controlled case numbers sufficiently until vaccine coverage was much higher. 

From October to December 2021, reduction in $R_\mathrm{eff}(t)$ as a result of increasing vaccine coverage was partially counteracted by the phased easing of restrictions and increasing contact rates. An ABC method was developed for inferring the value of the time-varying ``control function'' $C(t)$ in Eq. \eqref{bpm_equation} from data on new daily cases \citep{plank2022using}. Real-time estimation of $R_\mathrm{eff}(t)$ was also provided using the {\em EpiNow2} package, which uses a semi-mechanistic model based on the renewal equation, Eq. (\ref{eq:renewal}) \citep{abbott2020estimating,binny2022real}. This was suitable for short-term forecasts. However, the more mechanistic branching process model \citep{plank2022using} enabled projections of epidemic dynamics over a longer period of time, taking account of future vaccinations based on comprehensive data on booked vaccination appointments.

\subsubsection{Modelling the impact of Omicron}

When the B.1.1.529 (Omicron) variant began spreading globally in December 2021 \citep{viana2022rapid}, emerging estimates of Omicron's severity \citep{ward2022risk,nyberg2022comparative} and time-dependent vaccine effectiveness \citep{andrews2022covid} were combined with projected booster uptake to produce scenarios for an Omicron outbreak starting in New Zealand on different dates \citep{vattiato2022assessment}. 

Up to January 2022, New Zealand had recorded a cumulative total of only 3 cases per 1,000 people. Once Omicron began to spread in February 2022, the model was calibrated to real-time data to produce scenarios for the height and timing of the peak. These results were incorporated into dashboards for government policymakers and District Health Board planners. Scenarios were updated periodically as the model was refined to account for relaxation of social distancing behaviours,  changing age-structured contact rates, and reinfections \citep{vattiato2022modelling}. 

The model was extended to investigate the effect of a new variant and subsequently used to model New Zealand's BA.5 wave, which occurred in July 2022 \citep{lustig2023modelling}. Immunity parameters were calibrated using the model of \citep{khoury2021neutralizing,cromer2022neutralising} for the relationship between neutralising antibody titre and protection against different clinical endpoints, coupled with whole genome sequencing data on community cases \citep{esr2022prevalence}. Around June 2022, the stochastic branching process model was replaced with a deterministic ODE model with similar underlying structure. This less computationally expensive model allowed an improved method for calibrating model parameters using an ABC approach. At the time of writing in April 2023, the ODE model is still being updated and fitted to data, and used to evaluate intervention effectiveness and inform policy decisions \citep{datta2023modelling}.

\section{What has modelling delivered?}
\label{sec:what_has_modelling_delivered}

Mathematical modelling has delivered quantitative results and qualitative interpretations of raw epidemiological and clinical data streams to support government decision-making in response to the pandemic. These include:
\begin{itemize}
    \item Estimates of $R_\mathrm{eff}$ (and transmission potential) and short-term forecasts, which aim to predict new cases and hospital admissions in the next 1--4 weeks, supporting week-to-week monitoring and response to the highly dynamic situation.
    \item Medium-term projections, produced by calibrating a model to recent data and then running it forward in time under the assumption that key factors affecting transmission (e.g. contact patterns) do not change, or change only in certain instructive ways over the relevant time frame.
    \item Long-term scenario modelling, which is more hypothetical, and may be used to compare alternative policy choices or the outcomes of, for example, different levels of vaccine uptake, over a period of months or years. The qualitative outcomes of scenario modelling have supported decisions on overall strategic response as described in Sec. \ref{sec:modelling_tools}.
\end{itemize}

For all three use-cases of mathematical modelling listed above, the ability to incorporate mechanistic assumptions, for example about the size of the susceptible population, vaccination coverage and roll-out, and alternative future behavioural mixing patterns, is of great importance. However, short-term forecasting typically makes use of simpler models with fewer mechanistic assumptions, for both practical (e.g.,\ computational run-time) and theoretical (e.g.,\ statistical overfitting) reasons. Scenario modelling, with a focus on exploratory, more qualitative analyses, typically benefits from development of more detailed models capturing more of the intricacies of the transmission process.

Among many challenges faced by applied mathematicians and data analysts supporting the COVID-19 response, communication to both the public and decision makers on the distinction between these different uses of mathematical modelling has been prominent. For example, the outputs of scenario models, including both the simulation results and the reports describing them, were prone to misinterpretation. Interpretation must be nuanced due to the potential sensitivity to model assumptions and mechanisms that are either not included in the model or are subject to high uncertainty, such as behavioural change, antigenic evolution of the pathogen, or development and use of new pharmaceuticals. Scenario models were typically developed for, and of greatest value in, making relative comparisons among different scenarios (e.g.,\ alternative policy settings). However, there was a tendency for the results to be considered as absolute predictions, particularly in the media but also at times in decision-making contexts. These challenges are, of course, not unique to Australia and New Zealand, nor to epidemiological modelling, and have been well documented across a range of model-based scientific fields, for example climate modelling \citep{maslin2012climate}.

The types of results described above will be familiar to many readers and are typically quite visible: they are described in written reports and scientific publications. In short, they are the direct outputs of our mathematical and statistical work. However, there is a second, less tangible way in which the mathematical sciences have contributed to the COVID-19 response, by providing public health leaders and decision makers (i.e.,\ politicians) with insights derived from a distinct mathematical way of thinking. For example, an intuitive grasp of nonlinear dynamics and stochasticity can provide insights or perspectives that might not otherwise be made without this type of training.

As described earlier, this was evident in the first weeks of the pandemic, when ``simple'' modelling in Australia and New Zealand was able to demonstrate to decision makers that not only were our countries at grave risk, but that an early, definitive response could dramatically and effectively mitigate that risk. The findings were a natural, and to the applied mathematician entirely unsurprising, consequence of the nature of transmission dynamics, yet surprisingly and deeply influential to those charged with determining our nations' responses.

Later in the pandemic, an ability to reason on the probability of importation, and the consequences of re-establishment of transmission in different epidemiological contexts helped decision makers to quantify and respond to risk in an uncertain setting. As an example, when the trans-Tasman quarantine-free travel zone was first being considered in 2021, New Zealand modellers were asked to estimate how this would affect the risk of community outbreaks. At the time, prevalence in Australia was so low that the risk of a trans-Tasman traveller causing an outbreak in New Zealand was minimal. It would have been tempting not to bother with a formal model at all. However, the process of setting up a model led to the realisation that if the hotel quarantine slots previously taken by arrivals from Australia were re-allocated to other countries where infection rates were much higher, the number of infected arrivals would likely increase substantially. That would inevitably translate into a higher risk of a case leaking out into the community. This realisation, obvious in retrospect, was not something that had been considered at the time as the focus was on the risk posed by trans-Tasman travellers.

More generally, the approach to problem solving inherent in the mathematical and physical sciences, and the contribution it can make to decision making, is harder to articulate and evaluate compared to technical outputs such as those described in Sec. \ref{sec:modelling_tools}. While the capability is almost taken for granted by members of the mathematical and physical sciences communities, it is of fundamental importance in a range of application areas, and has proven invaluable during the COVID-19 pandemic. We suggest that as a community, we need to challenge ourselves to describe this skill set more clearly and more explicitly, to notice when we are using it, and to value it. We are of course not suggesting that mathematical models have the answers to all questions. Often, answers from a mathematical model disintegrate on contact with reality and should be quickly abandoned, although even then there are lessons: which assumption(s) invalidated the work and why? What could be changed or included to improve the model and its role in decision support? Furthermore, domain expertise (here in infectious disease epidemiology and public health), gained either directly through professional roles or through deep, multi-disciplinary collaboration, is absolutely crucial. Elegant, and perhaps even accurate mathematical models developed or presented without an understanding of the public health domain and context are unlikely to have impact. But when offered from an informed position, simple mathematical perspectives on complex problems are often illuminating and at times critical to informed decision making.

\section{Communication between modellers and stakeholders}
\label{sec:communication}

Communicating the purposes of modelling and how model outputs should and should not be interpreted has been a key challenge of working with policymakers and advisors in a fast-moving environment. Some stakeholders who were initially unfamiliar with modelling as a discipline had a tendency to view models as crystal balls, whereas other tended to dismiss any model as being too assumption-dependent or lacking sufficient complexity to be of value. The reality and appropriate use of model-based analyses lies in between these extremes, and we believe this has become much better understood by policymakers over time. At their core, models provide a way of representing a complicated system in a simplified way, thinking systematically about the consequences of a range of assumptions, and testing which of those assumptions are consistent with observed data.

It is the task of the modellers to find a balance between models that are too simplistic and miss essential features, and models that are overly complicated or include irrelevant variables or mechanisms. This balance will depend on the policy question the model is designed to address. In our experience, navigating this complex landscape benefits from strong and sustained stakeholder engagement and a rich bidirectional dialogue. Engagement and clear communication of the scope and purpose of modelling in general helps give stakeholders the confidence to interpret particular models, critically evaluate findings, engage in the process of model improvement and thereby incorporate findings from modelling into the decision-making process in a nuanced, robust way. If results can be transparently explained with reference to specific key assumptions, particularly if the results are unexpected or counter-intuitive, then decision makers can draw on their complementary expertise and knowledge from outside of the modelling domain. We argue that this enhances the value of modelling to the decision-making process. A corollary is that black-box type models or models developed and delivered in the absence of stakeholder input can be problematic, and are (in both our considered opinion and from experience) unlikely to be adopted by decision-makers and have impact.

Communicating limitations and uncertainty in model outputs is an ongoing challenge but an essential one: to avoid over-confidence in point estimates; to help understand the differences between the types of model output mentioned above; to help interpret outputs such as sensitivity analyses and confidence intervals; and to enable evaluation of model performance against subsequent observed data \citep{mccabe2021communicating}. Both frequentist and Bayesian statistical tools, such as maximum likelihood or approximate Bayesian computation, provide practical ways to fit models to observed data and produce associated uncertainty intervals on model outputs. Choice of appropriate summary statistics for the outputs of dynamic models is important to accurately convey the nature of the uncertainty. For example, these could include curve-based statistics as opposed to commonly used but often misleading fixed-time statistics \citep{juul2021fixed}, or probabilistic estimates of certain outcomes such as healthcare demand exceeding a specified threshold \citep{mccabe2021communicating}. Capturing uncertainty due to model mis-specification or unmodelled processes such as future new variants or unanticipated behavioural change is a bigger challenge. 

Distinguishing between situations with relatively high and low levels of uncertainty is important. For example, during the exponential growth or decline phase of a wave, the epidemic trajectory was relatively predictable and forecast uncertainty was typically low. In contrast, around the peak of a wave (or trough between waves), future epidemic activity is typically far more uncertain. While there may be a temptation to think that model outputs with very wide uncertainty intervals are not useful, the finding that future epidemic states cannot be predicted with confidence in some situations is itself valuable to decision-makers.

Communication is not just about mathematical modellers helping stakeholders to interpret model results. Two-way communication between policymakers and modellers about the policy questions that the model is being designed to address and the appropriate model design is equally important. Ideally, this is an interactive and iterative process so that models can be refined and updated as needed, for example according to data availability or updated policy objectives. Building awareness among policymakers of what models can and cannot deliver helps ensure modelling questions are designed appropriately and efficiently. Likewise rich two-way communication can help influence surveillance practices, leading to enhanced availability of high-quality data for more accurate modelling. Conversely, it is essential that modellers are aware of the biases and limitations of the data that are used as model inputs, which requires engagement with public health agencies gathering the data \citep{lipsitch2011improving}. 

Although modelling has mostly been very positively received in the policy arena in both Australia and New Zealand, it has at times been heavily criticised in other areas. Much of that criticism is valid: challenging particular modelling assumptions, mechanisms and trade offs that were not included in models, or inadequate communication of model interpretation. There are many aspects where modelling infectious disease dynamics for policy advice can be improved and this remains an active area of research \citep{bedson2021review,mccabe2021communicating}. At other times, criticism stemmed from a misunderstanding of the purposes and limitations of modelling, or simply a dislike of its conclusions or of the decisions which it informed. Many people working in this area, including both authors, have invested significant time and effort into communicating the purposes, interpretations and limitations of modelling to the general public via mainstream media \cite{medley2022aconsensus}. We believe this is an essential part of mathematical modelling for public policy advice, promoting transparency, trust and understanding of the modelling process, including what it can and cannot achieve. Communicating the value of mathematical and statistical training more broadly to policymakers and the general public is an ongoing challenge for our community.

\section{Future challenges and opportunities}

As Australia and New Zealand transition out of the acute phase of the COVID-19 pandemic, there are a number of opportunities, but also risks, for the mathematical sciences community and those involved in infectious disease modelling and response. First and foremost, the impact of mathematical modelling on the pandemic response has once again demonstrated the fundamental importance of our broad discipline to societal prosperity, health and well-being. While our observations on the value and practice of applied mathematical modelling have necessarily come from the epidemiological context, many of them apply equally to applications of mathematical modelling in other content and policy areas. 

Mathematical epidemiology provided insight into the pandemic, demonstrated how we could respond to it, and had a significant impact on government decision making and pandemic response strategy in both Australia and New Zealand. Of course, a mathematical model cannot tell society what to do: policy decisions rightly rest with elected politicians, often involve value judgements, and are typically made using a variety of sources of evidence. Nevertheless, mathematical modelling provided insight that was arguably not available through any other discipline base alone, including traditional public health and epidemiological training. Modelling and model-based analyses of data proved particularly valuable in fast-moving situations where decisions needed to be made before more complete information became available. Ultimately, the impact of modelling lay not in its ability to tell policymakers what to do (despite what some in society, the media and even political circles claimed at certain times), but in the fact that governments made decisions with a clearer understanding of the likely consequences of their choices than they would otherwise have had.

The success of applied mathematics and modelling during COVID-19 affords a huge opportunity for governments and the public health sector, in both academia and the public service, to set up new collaborations and training pathways in and with the mathematical sciences. Training in infectious disease epidemiology and public health could be fundamentally transformed with a renewed commitment to rigorous training in the mathematical sciences. We suggest this would need to begin in high school.

However, with such an opportunity, there is also risk that must be managed. When is the right time to ``specialise''? Neither of the authors of this work ever trained formally in epidemiology or public health, acquiring those skills in their professional post-PhD careers. If and when the skills of infectious disease modelling are taught routinely to those undertaking training outside of the mathematical sciences --- in say clinical, biomedical, health and life sciences --- do we risk losing the deeper expertise in applied mathematics, statistics and data science that were foundational to the teams and consortia delivering high-impact support to government during the pandemic? 

Our view is that this risk is manageable and the benefits that flow from strengthening and growing interdisciplinary collaborations between the mathematical sciences, epidemiology and public health, and other research disciplines and application areas are significant. There are enormous benefits to be had from training a future epidemiology and public health workforce in the core competencies of mathematical epidemiology. These include enhancing modelling capabilities outside academia, uncovering new research questions and developing application-oriented methodologies. This will benefit both the mathematical sciences and epidemiology and public health communities, building capacity that can be drawn on in a future emergency.

\section*{Acknowledgements}

The mathematical modelling and data analytics work described in this paper was developed and performed by a large number of people from across Australia and New Zealand, only a few of whom are mentioned by name in the manuscript. Rather than list them here and risk omissions, the authors direct the reader to the full co-authorship lists of referenced papers and the multiple consortium webpages referenced in the main text. MP acknowledges funding in support of the work described in this manuscript from Te P\=unaha Matatini, the New Zealand Ministry of Business, Innovation and Employment, Department of the Prime Minister and Cabinet, and Ministry of Health. MP acknowledges the role of the Ministry of Health, StatsNZ, and the Institute of Environmental Science and Research (ESR) in supplying data in support of this work. MP is grateful to members of the COVID-19 Modelling Government Steering Committee for helping to design the modelling questions that the work described in this manuscript addressed and for their role in communicating model outputs to decision makers. JM acknowledges funding support for the work described in this manuscript from the Australian Research Council, National Health and Medical Research Council, Defence Science Technology Group, United States Defence, and Australian Government Department of Health and Ageing. JM thanks fellow members of numerous national working groups and committees which contributed to the design, implementation and reporting of modelling and data-analytics studies described in this manuscript. The authors are grateful to two anonymous reviewers for comments on an earlier version of this manuscript.


\begin{thebibliography}{}

\bibitem[Abbott et~al., 2020]{abbott2020estimating}
Abbott, S., Hellewell, J., Thompson, R.~N., Sherratt, K., Gibbs, H.~P., Bosse,
  N.~I., Munday, J.~D., Meakin, S., Doughty, E.~L., Chun, J.~Y., Chan,
  Y.-W.~D., Finger, F., Campbell, P., Endo, A., Pearson, C. A.~B., Gimma, A.,
  Russell, T., {CMMID COVID modelling group}, Flashe, S., Kucharski, A.~J.,
  Eggo, R.~M., and Funk, S. (2020).
\newblock Estimating the time-varying reproduction number of {SARS-CoV-2} using
  national and subnational case counts.
\newblock {\em Wellcome Open Research}, 5:112.

\bibitem[Andrews et~al., 2022]{andrews2022covid}
Andrews, N., Stowe, J., Kirsebom, F., Toffa, S., Rickeard, T., Gallagher, E.,
  Gower, C., Kall, M., Groves, N., Oâ€™Connell, A.-M., Simons, D., Blomquist,
  P.~B., Zaidi, A., Nash, S., Iwani Binti Abdul~Aziz, N., Thelwall, S., Dabera,
  G., Myers, R., Amirthalingam, G., Gharbia, S., Barrett, J.~C., Elson, R.,
  Ladhani, S.~N., Ferguson, N., Zambon, M., Campbell, C. N.~J., Brown, K.,
  Hopkins, S., Chand, M., Ramsay, M., and Lopez~Bernal, J. (2022).
\newblock Covid-19 vaccine effectiveness against the {Omicron} {(B.1.1.529)}
  variant.
\newblock {\em New England Journal of Medicine}.

\bibitem[{Australian Government Department of Health and Aged Care},
  2022]{AUGov2022surveillanceplan}
{Australian Government Department of Health and Aged Care} (2022).
\newblock {A}ustralian {N}ational {D}isease {S}urveillance {P}lan for
  {COVID-19} ({V}ersion 3.0).
\newblock
  https://www.health.gov.au/resources/publications/australian-national-disease-surveillance-plan-for-covid-19.

\bibitem[Baker et~al., 2020a]{baker2020new}
Baker, M.~G., Kvalsvig, A., Verrall, A.~J., Telfar-Barnard, L., and Wilson, N.
  (2020a).
\newblock New {Z}ealand's elimination strategy for the {COVID}-19 pandemic and
  what is required to make it work.
\newblock {\em New Zealand Medical Journal}, 133(1512):10--14.

\bibitem[Baker et~al., 2020b]{baker2020successful}
Baker, M.~G., Wilson, N., and Anglemyer, A. (2020b).
\newblock Successful elimination of {Covid}-19 transmission in {New Zealand}.
\newblock {\em New England Journal of Medicine}, 383(8):e56.

\bibitem[Barnard et~al., 2022]{barnard2022modelling}
Barnard, R.~C., Davies, N.~G., Jit, M., and Edmunds, W.~J. (2022).
\newblock Modelling the medium-term dynamics of {SARS-CoV-2} transmission in
  {E}ngland in the {O}micron era.
\newblock {\em Nature Communications}, 13(1):4879.

\bibitem[Bedson et~al., 2021]{bedson2021review}
Bedson, J., Skrip, L.~A., Pedi, D., Abramowitz, S., Carter, S., Jalloh, M.~F.,
  Funk, S., Gobat, N., Giles-Vernick, T., Chowell, G., de~Almeida, J.~R.,
  Elessawi, R., Scarpino, S.~V., Hammond, R.~A., Briand, S., Epstein, J.~M.,
  H\'erbert-Dufresne, L., and Althouse, B.~M. (2021).
\newblock A review and agenda for integrated disease models including social
  and behavioural factors.
\newblock {\em Nature Human Behaviour}, 5:834--846.

\bibitem[Binny et~al., 2022]{binny2022real}
Binny, R.~N., Lustig, A., Hendy, S.~C., Maclaren, O.~J., Ridings, K.~M.,
  Vattiato, G., and Plank, M.~J. (2022).
\newblock Real-time estimation of the effective reproduction number of
  {SARS-CoV-2} in {A}otearoa {N}ew {Z}ealand.
\newblock {\em PeerJ}, 10:e14119.

\bibitem[Conway et~al., 2023]{conway2023covid}
Conway, E., Walker, C., Baker, C., Lydeamore, M., Ryan, G.~E., Campbell, T.,
  Miller, J.~C., Yeung, M., Kabashima, G., Wood, J., Rebuli, N., McCaw, J.~M.,
  McVernon, J., Golding, N., Price, D.~J., and Shearer, F.~M. (2023).
\newblock Covid-19 vaccine coverage targets to inform reopening plans in a low
  incidence setting.
\newblock {\em medRxiv}, doi.org/10.1101/2022.12.04.22282996.

\bibitem[Cromer et~al., 2022]{cromer2022neutralising}
Cromer, D., Steain, M., Reynaldi, A., Schlub, T.~E., Wheatley, A.~K., Juno,
  J.~A., Kent, S.~J., Triccas, J.~A., Khoury, D.~S., and Davenport, M.~P.
  (2022).
\newblock Neutralising antibody titres as predictors of protection against
  {SARS-CoV-2} variants and the impact of boosting: a meta-analysis.
\newblock {\em Lancet Microbe}, 3(1):e52--e61.

\bibitem[Datta et~al., 2023]{datta2023modelling}
Datta, S., Gilmour, J., Harvey, E., Maclaren, O.~J., O'Neale, D. R.~J.,
  Patten-Elliott, F., Plank, M.~J., Priest-Forsyth, E., Turnbull, S., Vattiato,
  G., and Wu, D. (2023).
\newblock Modelling the effect of changes to the {COVID-19} case isolation
  policy.
\newblock {\em Covid-19 Modelling Aotearoa},
  https://www.covid19modelling.ac.nz/modelling-the-effect-of-changes-to-the-covid-19-case-isolation-policy/.

\bibitem[Davies et~al., 2021a]{davies2021estimated}
Davies, N.~G., Abbott, S., Barnard, R.~C., Jarvis, C.~I., Kucharski, A.~J.,
  Munday, J.~D., Pearson, C.~A., Russell, T.~W., Tully, D.~C., Washburne,
  A.~D., Wenseleers, T., Gimma, A., Waites, W., Wong, K. L.~M., Van~Zandvoort,
  K., Silverman, J.~D., {CMMID COVID-19 Working Group}, {COVID-19 Genomics UK
  Consortium}, Diaz-Ordaz, K., Keogh, R., Eggo, R.~M., Funk, S., Jit, M.,
  Atkins, W.~E., and Edmunds, W.~J. (2021a).
\newblock Estimated transmissibility and impact of {SARS-CoV-2} lineage
  {B.1.1.7} in {E}ngland.
\newblock {\em Science}, 372(6538):eabg3055.

\bibitem[Davies et~al., 2021b]{davies2021increased}
Davies, N.~G., Jarvis, C.~I., Edmunds, W.~J., Jewell, N.~P., Diaz-Ordaz, K.,
  and Keogh, R.~H. (2021b).
\newblock Increased mortality in community-tested cases of {SARS-CoV-2} lineage
  {B.1.1.7}.
\newblock {\em Nature}, 593(7858):270--274.

\bibitem[Davies et~al., 2020]{davies2020age}
Davies, N.~G., Klepac, P., Liu, Y., Prem, K., Jit, M., and Eggo, R.~M. (2020).
\newblock Age-dependent effects in the transmission and control of {COVID}-19
  epidemics.
\newblock {\em Nature Medicine}, 26(8):1205--1211.

\bibitem[Douglas et~al., 2021]{douglas2021real}
Douglas, J., Geoghegan, J.~L., Hadfield, J., Bouckaert, R., Storey, M., Ren,
  X., de~Ligt, J., French, N., and Welch, D. (2021).
\newblock Real-time genomics for tracking severe acute respiratory syndrome
  coronavirus 2 border incursions after virus elimination, {New Zealand}.
\newblock {\em Emerging Infectious Diseases}, 27(9):2361.

\bibitem[Du et~al., 2020]{du2020serial}
Du, Z., Xu, X., Wu, Y., Wang, L., Cowling, B.~J., and Meyers, L.~A. (2020).
\newblock Serial interval of {COVID-19} among publicly reported confirmed
  cases.
\newblock {\em Emerging Infectious Diseases}, 26:1341.

\bibitem[Dyson et~al., 2021]{dyson2021possible}
Dyson, L., Hill, E.~M., Moore, S., Curran-Sebastian, J., Tildesley, M.~J.,
  Lythgoe, K.~A., House, T., Pellis, L., and Keeling, M.~J. (2021).
\newblock Possible future waves of {SARS-CoV-2} infection generated by variants
  of concern with a range of characteristics.
\newblock {\em Nature Communications}, 12(1):5730.

\bibitem[ESR, 2022]{esr2022prevalence}
ESR (2022).
\newblock Prevelence of {SARS-CoV-2} variants of concern in {A}oteoroa {N}ew
  {Z}ealand.
\newblock https://github.com/ESR-NZ/nz-sars-cov2-variants.

\bibitem[Flaxman et~al., 2020]{flaxman2020estimating}
Flaxman, S., Mishra, S., Gandy, A., Unwin, H. J.~T., Mellan, T.~A., Coupland,
  H., Whittaker, C., Zhu, H., Berah, T., Eaton, J.~W., Monod, M., {Imperial
  College COVID-19 Response Team}, Ghani, A.~C., Donnelly, C.~A., Riley, S.,
  Vollmer, M. A.~C., Ferguson, N.~M., Okell, L.~C., and Bhatt, S. (2020).
\newblock Estimating the effects of non-pharmaceutical interventions on
  {COVID}-19 in {E}urope.
\newblock {\em Nature}, 584(7820):257--261.

\bibitem[Gilmour et~al., 2021a]{gilmour2021modelling}
Gilmour, J., Harvey, E., Looker, J., Mackenzie, F., Maclaren, O., O'Neale, D.,
  Patten-Elliott, F., Trent, J., Turnbull, S., and Wu, D. (2021a).
\newblock Modelling estimates of expected size: the {A}ugust 2021 {COVID}-19
  outbreak in {A}otearoa.
\newblock {\em Covid-19 Modelling Aotearoa},
  https://www.covid19modelling.ac.nz/contagion-network-modelling-in-the-first-weeks-of-the-august-2021-outbreak/.

\bibitem[Gilmour et~al., 2021b]{gilmour2021inter}
Gilmour, J., Harvey, E., Looker, J., Mackenzie, F., O'Neale, D., and Turnbull,
  S. (2021b).
\newblock Inter-regional movement and contagion risk analysis {A}ugust 2021.
\newblock {\em Covid-19 Modelling Aotearoa},
  https://www.covid19modelling.ac.nz/inter-regional-movement-and-contagion-risk/.

\bibitem[Golding et~al., 2023]{golding2023modelling}
Golding, N., Price, D.~J., Ryan, G., McVernon, J., McCaw, J.~M., and Shearer,
  F.~M. (2023).
\newblock A modelling approach to estimate the transmissibility of {SARS-CoV-2}
  during periods of high, low, and zero case incidence.
\newblock {\em eLife}, 12:e78089.

\bibitem[Golding et~al., 2020a]{IDDU2020techreport2}
Golding, N., Price, D.~J., Shearer, F.~M., Moss, R., Meehan, M.~T., McBryde,
  E., Dawson, P., McVernon, J., and McCaw, J.~M. (2020a).
\newblock Estimating temporal variation in transmission of {COVID-19} and
  adherence to social distancing measures in {A}ustralia: {T}echnical {R}eport
  5th {M}ay 2020.
\newblock IDDU Technical Report:
  https://mspgh.unimelb.edu.au/\_\_data/assets/pdf\_file/0005/4230644/2020-05-05-Technical-report-public-release.pdf.

\bibitem[Golding et~al., 2020b]{IDDU2020techreport3}
Golding, N., Shearer, F.~M., Moss, R., Dawson, P., Gibbs, L., Alisic, E.,
  McVernon, J., Price, D.~J., and McCaw, J.~M. (2020b).
\newblock Estimating temporal variation in transmission of {COVID-19} and
  adherence to social distancing measures in {A}ustralia: {T}echnical {R}eport
  15th {M}ay 2020.
\newblock IDDU Technical Report:
  https://mspgh.unimelb.edu.au/\_\_data/assets/pdf\_file/0006/4230645/2020-05-15-Technical\_report-public-release.pdf.

\bibitem[Golding et~al., 2021]{IDDU2021techreport}
Golding, N., Shearer, F.~M., Moss, R., Dawson, P., Liu, D., Ross, J.~V.,
  Hyndman, R., Montero-Manso, P., Ryan, G., South, T., McVernon, J., Price,
  D.~J., and McCaw, J.~M. (2021).
\newblock Situational assessment of {COVID-19} in {A}ustralia: {T}echnical
  {R}eport 15th {M}arch 2021 (released 28 {M}ay 2021).
\newblock IDDU Technical Report:
  https://mspgh.unimelb.edu.au/\_\_data/assets/pdf\_file/0004/4230643/2021-03-15-Technical-report-public-release.pdf.

\bibitem[Golding et~al., 2020c]{IDDU2020techreport4}
Golding, N., Shearer, F.~M., Moss, R., Dawson, P., Liu, D., Ross, J.~V.,
  Hyndman, R., Zachreson, C., Geard, N., McVernon, J., Price, D.~J., and McCaw,
  J.~M. (2020c).
\newblock Estimating the temporal variation in transmission of {SARS-CoV-2} and
  physical distancing behaviour in {A}ustralia: {T}echnical {R}eport 17th
  {J}uly 2020.
\newblock IDDU Technical Report:i
  https://mspgh.unimelb.edu.au/\_\_data/assets/pdf\_file/0009/4231188/2020-07-17-Technical\_report-public-release.pdf.

\bibitem[Grout et~al., 2021]{grout2021failures}
Grout, L., Katar, A., Ait~Ouakrim, D., Summers, J.~A., Kvalsvig, A., Baker,
  M.~G., Blakely, T., and Wilson, N. (2021).
\newblock Failures of quarantine systems for preventing {COVID}-19 outbreaks in
  {Australia} and {New Zealand}.
\newblock {\em Medical Journal of Australia}, 215(7):320--324.

\bibitem[Harvey et~al., 2020]{harvey2020network}
Harvey, E., Maclaren, O., O'Neale, D., Ortiz-Cervantes, A., Patten-Elliott, F.,
  Turnbull, S., {Vasques Filho}, D., and Wu, D. (2020).
\newblock Network-based simulations of re-emergence and spread of {COVID}-19 in
  {A}otearoa {N}ew {Z}ealand.
\newblock {\em Covid-19 Modelling Aotearoa},
  www.covid19modelling.ac.nz/simulations-of-re-emergence-and-spread/.

\bibitem[Harvey et~al., 2021]{harvey2021contagion}
Harvey, E., Maclaren, O., O'Neale, D., Patten-Elliott, F., Turnbull, S., and
  Wu, D. (2021).
\newblock Contagion network modelling of effectiveness for a range of
  non-pharmaceutical interventions for {COVID}-19 elimination in {A}otearoa
  {N}ew {Z}ealand.
\newblock {\em Covid-19 Modelling Aotearoa},
  https://www.covid19modelling.ac.nz/network-modelling-trilogy/.

\bibitem[Hendy, 2022]{hendy2022integrating}
Hendy, S. (2022).
\newblock Integrating science into policy: experiences during the pandemic.
\newblock {\em Policy Quarterly}, 18(1):38--43.

\bibitem[Hendy et~al., 2021]{hendy2021mathematical}
Hendy, S., Steyn, N., James, A., Plank, M.~J., Hannah, K., Binny, R.~N., and
  Lustig, A. (2021).
\newblock Mathematical modelling to inform {New Zealand}â€™s {COVID}-19
  response.
\newblock {\em Journal of the Royal Society of New Zealand},
  51(sup1):S86--S106.

\bibitem[Herrera-Esposito and de~Los~Campos, 2022]{herrera2022age}
Herrera-Esposito, D. and de~Los~Campos, G. (2022).
\newblock Age-specific rate of severe and critical {SARS-CoV-2} infections
  estimated with multi-country seroprevalence studies.
\newblock {\em BMC Infectious Diseases}, 22(1):311.

\bibitem[Hsiang et~al., 2020]{hsiang2020effect}
Hsiang, S., Allen, D., Annan-Phan, S., Bell, K., Bolliger, I., Chong, T.,
  Druckenmiller, H., Huang, L.~Y., Hultgren, A., Krasovich, E., et~al. (2020).
\newblock The effect of large-scale anti-contagion policies on the {COVID}-19
  pandemic.
\newblock {\em Nature}, 584(7820):262--267.

\bibitem[James et~al., 2020]{james2020suppression}
James, A., Hendy, S.~C., Plank, M.~J., and Steyn, N. (2020).
\newblock Suppression and mitigation strategies for control of {COVID}-19 in
  {New Zealand}.
\newblock {\em medRxiv}, 2020.03.26.20044677.

\bibitem[James et~al., 2021]{james2021successful}
James, A., Plank, M.~J., Hendy, S., Binny, R., Lustig, A., Steyn, N., Nesdale,
  A., and Verrall, A. (2021).
\newblock Successful contact tracing systems for {COVID}-19 rely on effective
  quarantine and isolation.
\newblock {\em PLoS ONE}, 16(6):e0252499.

\bibitem[Jelley et~al., 2022]{jelley2022genomic}
Jelley, L., Douglas, J., Ren, X., Winter, D., McNeill, A., Huang, S., French,
  N., Welch, D., Hadfield, J., de~Ligt, J., and Geoghegan, J.~L. (2022).
\newblock Genomic epidemiology of {D}elta {SARS-CoV-2} during transition from
  elimination to suppression in {A}otearoa {N}ew {Z}ealand.
\newblock {\em Nature Communications}, 13(1):4035.

\bibitem[Juul et~al., 2021]{juul2021fixed}
Juul, J.~L., Gr{\ae}sb{\o}ll, K., Christiansen, L.~E., and Lehmann, S. (2021).
\newblock Fixed-time descriptive statistics underestimate extremes of epidemic
  curve ensembles.
\newblock {\em Nature Physics}, 17(1):5--8.

\bibitem[Keeling et~al., 2021a]{keeling2021short}
Keeling, M.~J., Brooks-Pollock, E., Challen, R., Danon, L., Dyson, L., Gog,
  J.~R., Guzm{\'a}n~Rinc{\'o}n, L., Hill, E.~M., Pellis, L., Read, J.~M., and
  Tildesley, M.~J. (2021a).
\newblock Short-term projections based on early {O}micron variant dynamics in
  {E}ngland.
\newblock {\em MedRxiv}, pages 2021--12.

\bibitem[Keeling et~al., 2022]{keeling2022comparison}
Keeling, M.~J., Dyson, L., Tildesley, M.~J., Hill, E.~M., and Moore, S. (2022).
\newblock Comparison of the 2021 {COVID}-19 roadmap projections against public
  health data in {E}ngland.
\newblock {\em Nature Communications}, 13(1):1--19.

\bibitem[Keeling et~al., 2021b]{keeling2021waning}
Keeling, M.~J., Thomas, A., Hill, E.~M., Thompson, R.~N., Dyson, L., Tildesley,
  M.~J., and Moore, S. (2021b).
\newblock Waning, boosting and a path to endemicity for {SARS-CoV-2}.
\newblock {\em medRxiv}, doi.org/10.1101/2021.11.05.21265977.

\bibitem[Khoury et~al., 2021]{khoury2021neutralizing}
Khoury, D.~S., Cromer, D., Reynaldi, A., Schlub, T.~E., Wheatley, A.~K., Juno,
  J.~A., Subbarao, K., Kent, S.~J., Triccas, J.~A., and Davenport, M.~P.
  (2021).
\newblock Neutralizing antibody levels are highly predictive of immune
  protection from symptomatic {SARS-CoV-2} infection.
\newblock {\em Nature Medicine}, 27(7):1205--1211.

\bibitem[Kucharski et~al., 2020a]{kucharski2020therole}
Kucharski, A.~J., Funk, S., and Eggo, R.~M. (2020a).
\newblock The {COVID-19} response illustrates that traditional academic reward
  structures and metrics do not reflect crucial contributions to modern
  science.
\newblock {\em PLOS Biology}, 18(10):1--3.

\bibitem[Kucharski et~al., 2020b]{kucharski2020early}
Kucharski, A.~J., Russell, T.~W., Diamond, C., Liu, Y., Edmunds, J., Funk, S.,
  Eggo, R.~M., Sun, F., Jit, M., Munday, J.~D., Davies, N., Gimma, A., {van
  Zandvoort}, K., Gibbs, H., Hellewell, J., Jarvis, C.~I., Clifford, S.,
  Quilty, B.~J., Bosse, N.~I., Abbott, S., Klepac, P., and Flasche, S. (2020b).
\newblock Early dynamics of transmission and control of {COVID-19}: a
  mathematical modelling study.
\newblock {\em Lancet Infectious Diseases}, 20(5):553--558.

\bibitem[Li et~al., 2020]{li2020early}
Li, Q., Guan, X., Wu, P., Wang, X., Zhou, L., Tong, Y., Ren, R., Leung, K.~S.,
  Lau, E.~H., Wong, J.~Y., Xing, X., Xiang, N., Wu, Y., Li, C., Chen, Q., Li,
  D., Liu, T., Zhao, J., Liu, M., Tu, W., Chen, C., Jin, L., Yang, R., Wang,
  Q., Zhou, S., Wang, R., Liu, H., Luo, Y., Liu, Y., Shao, G., Li, H., Tao, Z.,
  Yang, Y., Deng, Z., Liu, B., Ma, Z., Zhang, Y., Shi, G., Lam, T.~T., Wu,
  J.~T., Gao, G.~F., Cowling, B.~J., Yang, B., Leung, G.~M., and Feng, Z.
  (2020).
\newblock Early transmission dynamics in {W}uhan, {C}hina, of novel
  coronavirusâ€“infected pneumonia.
\newblock {\em New England Journal of Medicine}, 382(13):1199--1207.

\bibitem[Lipsitch et~al., 2011]{lipsitch2011improving}
Lipsitch, M., Finelli, L., Heffernan, R.~T., Leung, G.~M., Redd, S.~C., and
  {2009 H1N1 Surveillance Group} (2011).
\newblock Improving the evidence base for decision making during a pandemic:
  the example of 2009 influenza {A/H1N1}.
\newblock {\em Biosecurity and Bioterrorism}, 9(2):89--115.

\bibitem[Lloyd-Smith et~al., 2005]{lloyd2005superspreading}
Lloyd-Smith, J.~O., Schreiber, S.~J., Kopp, P.~E., and Getz, W.~M. (2005).
\newblock Superspreading and the effect of individual variation on disease
  emergence.
\newblock {\em Nature}, 438(7066):355--359.

\bibitem[Lustig et~al., 2023]{lustig2023modelling}
Lustig, A., Vattiato, G., Maclaren, O., Watson, L.~M., Datta, S., and Plank,
  M.~J. (2023).
\newblock Modelling the impact of the {O}micron {BA.5} subvariant in {N}ew
  {Z}ealand.
\newblock {\em Journal of the Royal Society Interface}, 20(199):20220698.

\bibitem[Maslin and Austin, 2012]{maslin2012climate}
Maslin, M. and Austin, P. (2012).
\newblock Climate models at their limit?
\newblock {\em Nature}, 486(7402):183--184.

\bibitem[McCabe et~al., 2021]{mccabe2021communicating}
McCabe, R., Kont, M.~D., Schmit, N., Whittaker, C., L{\o}chen, A., Walker,
  P.~G., Ghani, A.~C., Ferguson, N.~M., White, P.~J., Donnelly, C.~A., and
  Watson, O.~J. (2021).
\newblock Communicating uncertainty in epidemic models.
\newblock {\em Epidemics}, 37:100520.

\bibitem[McCaw and McVernon, 2007]{mccaw2007prophylaxis}
McCaw, J.~M. and McVernon, J. (2007).
\newblock Prophylaxis or treatment? {O}ptimal use of an antiviral stockpile
  during an influenza pandemic.
\newblock {\em Math. Biosci.}, 209:336--360.

\bibitem[McCaw et~al., 2022]{IDDU2022techreport}
McCaw, J.~M., Moss, R., Price, D.~J., Shearer, F.~M., Tobin, R., Golding, N.,
  Hao, T., Ryan, G., Adekunle, A., Dawson, P., Teo, M., Morris, D., Ross,
  J.~V., South, T., Hyndman, R., Lydeamore, M., O'Hara-Wild, M., and Wood, J.
  (2022).
\newblock Situational assessment of {COVID-19} in {A}ustralia: {T}echnical
  {R}eport 22nd {M}ay 2022 (released 12 {A}ugust 2022).
\newblock IDDU Technical Report:
  https://mspgh.unimelb.edu.au/\_\_data/assets/pdf\_file/0003/4256103/2022-05-22-Technical-report-public-release.pdf.

\bibitem[Medley, 2022]{medley2022aconsensus}
Medley, G.~F. (2022).
\newblock A consensus of evidence: {T}he role of {SPI-M-O} in the {UK}
  {COVID-19} response.
\newblock {\em Advances in Biological Regulation}, 86:100918.

\bibitem[Moore et~al., 2021]{moore2021vaccination}
Moore, S., Hill, E.~M., Tildesley, M.~J., Dyson, L., and Keeling, M.~J. (2021).
\newblock Vaccination and non-pharmaceutical interventions for {COVID}-19: a
  mathematical modelling study.
\newblock {\em Lancet Infectious Diseases}, 21(6):793--802.

\bibitem[Moss et~al., 2018]{moss2018epidemic}
Moss, R., Fielding, J.~E., Franklin, L.~J., Stephens, N., McVernon, J., Dawson,
  P., and McCaw, J.~M. (2018).
\newblock Epidemic forecasts as a tool for public health: interpretation and
  (re)calibration.
\newblock {\em Australian and New Zealand Journal of Public Health}, 42:69--76.

\bibitem[Moss et~al., 2022]{moss2022forecast}
Moss, R., Price, D.~J., Golding, N., Dawson, P., McVernon, J., Hyndman, R.~J.,
  Shearer, F.~M., and McCaw, J.~M. (2022).
\newblock Forecasting {COVID-19} activity in {A}ustralia to support pandemic
  response: {M}ay to {O}ctober 2020.
\newblock {\em medRxiv}, 2022.08.04.22278391.

\bibitem[Moss et~al., 2020]{moss2020coronavirus}
Moss, R., Wood, J., Brown, D., Shearer, F.~M., Black, A.~J., Glass, K., Cheng,
  A.~C., McCaw, J.~M., and McVernon, J. (2020).
\newblock Coronavirus disease model to inform transmission-reducing measures
  and health system preparedness, {A}ustralia.
\newblock {\em Emerging Infectious Diseases}, 26:2844--2853.

\bibitem[Moss et~al., 2016]{moss2016forecasting}
Moss, R., Zarebski, A., Dawson, P., and McCaw, J.~M. (2016).
\newblock Forecasting influenza outbreak dynamics in {M}elbourne from internet
  search query surveillance data.
\newblock {\em Influenza and Other Respiratory Viruses}.

\bibitem[Moss et~al., 2017]{moss2017retrospective}
Moss, R., Zarebski, A., Dawson, P., and McCaw, J.~M. (2017).
\newblock Retrospective forecasting of the 2010-2014 {M}elbourne influenza
  seasons using multiple surveillance systems.
\newblock {\em Epidemiology and Infection}, 145:156--169.

\bibitem[Moss et~al., 2019a]{moss2019accounting}
Moss, R., Zarebski, A.~E., Carlson, S.~J., and McCaw, J.~M. (2019a).
\newblock Accounting for healthcare-seeking behaviours and testing practices in
  real-time influenza forecasts.
\newblock {\em Tropical Medicine and Infectious Disease}, 4(1).

\bibitem[Moss et~al., 2019b]{moss2019anatomy}
Moss, R., Zarebski, A.~E., Dawson, P., Franklin, L.~J., Birrell, F.~A., and
  McCaw, J.~M. (2019b).
\newblock Anatomy of a seasonal influenza epidemic forecast.
\newblock {\em Communicable Diseases Intelligence}, 43.

\bibitem[Mossong et~al., 2008]{mossong2008social}
Mossong, J., Hens, N., Jit, M., Beutels, P., Auranen, K., Mikolajczyk, R.,
  Massari, M., Salmaso, S., Tomba, G.~S., Wallinga, J., et~al. (2008).
\newblock Social contacts and mixing patterns relevant to the spread of
  infectious diseases.
\newblock {\em PLoS Medicine}, 5(3):e74.

\bibitem[{New Zealand Government}, 2020a]{nzgovt2020covid}
{New Zealand Government} (2020a).
\newblock {COVID-19} resurgence: Improving public health measures at alert
  level 1.
\newblock Office of the Minister for COVID-19 Response:
  covid19.govt.nz/assets/Proactive-Releases/Alert-levels-and-restrictions/22-Dec-2021/Paper-CP1-16112021-COVID-19-Resurgence-Improving-Public-Health-Measures-at-Alert-Level-1.pdf.

\bibitem[{New Zealand Government}, 2020b]{nzgovt2020AL1}
{New Zealand Government} (2020b).
\newblock New {Z}ealand moves to {A}lert {L}evel 1.
\newblock https://www.beehive.govt.nz/release/new-zealand-moves-alert-level-1.

\bibitem[Nyberg et~al., 2022]{nyberg2022comparative}
Nyberg, T., Ferguson, N.~M., Nash, S.~G., Webster, H.~H., Flaxman, S., Andrews,
  N., Hinsley, W., Bernal, J.~L., Kall, M., Bhatt, S., Blomquist, P., Zaidi,
  A., Volz, E., Aziz, N.~A., Harman, K., Funk, S., Abbott, S., {COVID-19
  Genomics UK consortium}, Hope, R., Charlett, A., Chand, M., Ghani, A.~C.,
  Seaman, S.~R., Dabrera, G., De~Angelis, D., Presanis, A.~M., and Thelwall, S.
  (2022).
\newblock Comparative analysis of the risks of hospitalisation and death
  associated with {SARS-CoV-2} {Omicron} ({B.1.1.529}) and {Delta}
  ({B.1.617.2}) variants in {England}: a cohort study.
\newblock {\em Lancet}, 399:1303--1312.

\bibitem[{Peter Doherty Institute}, 2021]{Doherty2021natplanwebsite}
{Peter Doherty Institute} (2021).
\newblock {D}oherty {I}nstitute {COVID-19} modelling to support the {N}ational
  {P}lan to transition {A}ustralia's {N}ational {COVID} {R}esponse.
\newblock
  https://www.doherty.edu.au/our-work/institute-themes/viral-infectious-diseases/covid-19/covid-19-modelling/modelling.

\bibitem[Plank et~al., 2021]{plank2021vaccination}
Plank, M.~J., Binny, R.~N., Hendy, S.~C., Lustig, A., and Ridings, K. (2021).
\newblock Vaccination and testing of the border workforce for {COVID}-19 and
  risk of community outbreaks: a modelling study.
\newblock {\em Royal Society Open Science}, 8(9):210686.

\bibitem[Plank et~al., 2022]{plank2022using}
Plank, M.~J., Hendy, S.~C., Binny, R.~N., Vattiato, G., Lustig, A., and
  Maclaren, O.~J. (2022).
\newblock Using mechanistic model-based inference to understand and project
  epidemic dynamics with time-varying contact and vaccination rates.
\newblock {\em Scientific Reports}, 12(1):20451.

\bibitem[Prem et~al., 2017]{prem2017projecting}
Prem, K., Cook, A.~R., and Jit, M. (2017).
\newblock Projecting social contact matrices in 152 countries using contact
  surveys and demographic data.
\newblock {\em PLoS Computational Biology}, 13(9):e1005697.

\bibitem[Price et~al., 2020a]{IDDU2020techreport1}
Price, D.~J., Shearer, F.~M., Meehan, M., McBryde, E., Golding, N., McVernon,
  J., and McCaw, J.~M. (2020a).
\newblock Estimating the case detection rate and temporal variation in
  transmission of {COVID-19} in {A}ustralia: {T}echnical {R}eport 14th {A}pril
  2020.
\newblock IDDU Technical Report:
  https://mspgh.unimelb.edu.au/\_\_data/assets/pdf\_file/0007/4230646/2020-04-14-Technical-report-public-release.pdf.

\bibitem[Price et~al., 2020b]{price2020early}
Price, D.~J., Shearer, F.~M., Meehan, M.~T., McBryde, E., Moss, R., Golding,
  N., Conway, E.~J., Dawson, P., Cromer, D., Wood, J., Abbott, S., McVernon,
  J., and McCaw, J.~M. (2020b).
\newblock Early analysis of the {A}ustralian {COVID}-19 epidemic.
\newblock {\em eLife}, 9:e58785.

\bibitem[Ryan et~al., 2022]{ryan2022estimating}
Ryan, G.~E., Shearer, F.~M., McCaw, J.~M., McVernon, J., and Golding, N.
  (2022).
\newblock Estimating measures to reduce the transmission of {SARS-CoV-2} in
  {A}ustralia to guide a `national plan' to reopening.
\newblock {\em medRxiv}, doi.org/10.1101/2022.12.15.22282869.

\bibitem[Shearer et~al., 2023]{shearer2023estimating}
Shearer, F.~M., McCaw, J.~M., Ryan, G.~E., Hao, T., Tierney, N., Lydeamore, M.,
  Ellis, S., Ward, K., Wood, J., McVernon, J., and Golding, N. (2023).
\newblock Estimating the impact of test-trace-isolate-quarantine systems on
  {SARS-CoV-2} transmission in {A}ustralia.
\newblock {\em medRxiv}, 10.1101/2023.01.10.23284209.

\bibitem[Shearer et~al., 2020]{shearer2020infectious}
Shearer, F.~M., Moss, R., McVernon, J., Ross, J.~V., and McCaw, J.~M. (2020).
\newblock Infectious disease pandemic planning and response: Incorporating
  decision analysis.
\newblock {\em PLOS Medicine}, 17:e1003018.

\bibitem[Shearer et~al., 2021]{shearer2021development}
Shearer, F.~M., Moss, R., Price, D.~J., Zarebski, A.~E., Ballard, P.~G.,
  McVernon, J., Ross, J.~V., and McCaw, J.~M. (2021).
\newblock Development of an influenza pandemic decision support tool linking
  situational analytics to national response policy.
\newblock {\em Epidemics}, 36:100478.

\bibitem[Shearer et~al., 2022]{shearer2022rapid}
Shearer, F.~M., Walker, J., Tellioglu, N., McCaw, J.~M., McVernon, J., Black,
  A., and Geard, N. (2022).
\newblock Rapid assessment of the risk of {SARS-CoV-2} importation: Case study
  and lessons learned.
\newblock {\em Epidemics}, 38:100549.

\bibitem[Sherratt et~al., 2023]{sherratt2023improving}
Sherratt, K., Carnegie, A.~C., Kucharski, A., Cori, A., Pearson, C.~A., Jarvis,
  C.~I., Overton, C., Weston, D., Hill, E.~M., Knock, E., Fearon, E.,
  Nightingale, E., Hellewell, J., Edmunds, W.~J., Arenas, J.~V., Prem, K., Pi,
  L., Baguelin, M., Kendall, M., Ferguson, N., Davies, N., Eggo, R.~M., van
  Elsland, S., Russell, T., Funk, S., Liu, Y., and Abbott, S. (2023).
\newblock Improving modelling for epidemic responses: reflections from members
  of the uk infectious disease modelling community on their experiences during
  the {COVID}-19 pandemic.
\newblock {\em bioRxiv}, 10.1101/2023.06.12.544667.

\bibitem[StatsNZ, 2022]{statsnz_idi}
StatsNZ (2022).
\newblock Integrated data infrastructure.
\newblock
  https://www.stats.govt.nz/integrated-data/integrated-data-infrastructure/.

\bibitem[Steyn et~al., 2020]{steyn2020estimated}
Steyn, N., Binny, R., Hannah, K., Hendy, S., James, A., Kukutai, T., Lustig,
  A., McLeod, M., Plank, M.~J., Ridings, K., and Sporle, A. (2020).
\newblock Estimated inequities in {COVID}-19 infection fatality rates by
  ethnicity for {A}otearoa {N}ew {Z}ealand.
\newblock {\em New Zealand Medical Journal}, 133:28--39.

\bibitem[Steyn et~al., 2021a]{steyn2021maori}
Steyn, N., Binny, R.~N., Hannah, K., Hendy, S., James, A., Lustig, A., Ridings,
  K., Plank, M.~J., and Sporle, A. (2021a).
\newblock M{\=a}ori and {Pacific} people in {New Zealand} have higher risk of
  hospitalisation for {COVID}-19.
\newblock {\em New Zealand Medical Journal}, 134:28--43.

\bibitem[Steyn et~al., 2022a]{steyn2022effect}
Steyn, N., Lustig, A., Hendy, S.~C., Binny, R.~N., and Plank, M.~J. (2022a).
\newblock Effect of vaccination, border testing, and quarantine requirements on
  the risk of {COVID}-19 in {N}ew {Z}ealand: A modelling study.
\newblock {\em Infectious Disease Modelling}, 7(1):184--198.

\bibitem[Steyn et~al., 2022b]{steyn2022covid}
Steyn, N., Plank, M.~J., Binny, R.~N., Hendy, S.~C., Lustig, A., and Ridings,
  K. (2022b).
\newblock A {COVID}-19 vaccination model for {Aotearoa New Zealand}.
\newblock {\em Scientific Reports}, 12(1):1--11.

\bibitem[Steyn et~al., 2021b]{steyn2021managing}
Steyn, N., Plank, M.~J., James, A., Binny, R.~N., Hendy, S.~C., and Lustig, A.
  (2021b).
\newblock Managing the risk of a {COVID}-19 outbreak from border arrivals.
\newblock {\em Journal of the Royal Society Interface}, 18(177):20210063.

\bibitem[Tobin et~al., 2023]{tobin2023real-time}
Tobin, R.~J., Wood, J.~G., Jayasundara, D., Sara, G., Walker, C.~R., Martin,
  G.~E., McCaw, J.~M., Shearer, F.~M., and Price, D.~J. (2023).
\newblock {Real-time analysis of hospital length of stay in a mixed SARS-CoV-2
  Omicron and Delta epidemic in New South Wales, Australia}.
\newblock {\em BMC Infectious Diseases}, 23(1).

\bibitem[Vattiatio et~al., 2022]{vattiato2022modelling}
Vattiatio, G., Lustig, A., Maclaren, O.~J., and Plank, M.~J. (2022).
\newblock Modelling the dynamics of infection, waning of immunity and
  re-infection with the {O}micron variant of {SARS-CoV-2} in {A}otearoa {N}ew
  {Z}ealand.
\newblock {\em Epidemics}, 41:100657.

\bibitem[Vattiato et~al., 2023]{vattiato2023modelling}
Vattiato, G., Lustig, A., Maclaren, O., Binny, R.~N., Hendy, S.~C., Harvey, E.,
  O'Neale, D., and Plank, M.~J. (2023).
\newblock Modelling {A}otearoa {N}ew {Z}ealand's {COVID}-19 protection
  framework and the transition away from the elimination strategy.
\newblock {\em Royal Society Open Science}, 10(2):220766.

\bibitem[Vattiato et~al., 2022]{vattiato2022assessment}
Vattiato, G., Maclaren, O., Lustig, A., Binny, R.~N., Hendy, S.~C., and Plank,
  M.~J. (2022).
\newblock An assessment of the potential impact of the {Omicron} variant of
  {SARS-CoV-2} in {Aotearoa New Zealand}.
\newblock {\em Infectious Disease Modelling}, 7:94--105.

\bibitem[Verity et~al., 2020]{verity2020estimates}
Verity, R., Okell, L.~C., Dorigatti, I., Winskill, P., Whittaker, C., Imai, N.,
  Cuomo-Dannenburg, G., Thompson, H., Walker, P. G.~T., Fu, H., Dighe, A.,
  Griffin, J.~T., Baguelin, M., Bhatia, S., Boonyasiri, A., Cori, A.,
  Cucunub\'a, Z., FitzJohn, R., Gaythorpe, K., Green, W., Hamlet, A., Hinsley,
  W., Laydon, D., Nedjati-Gilani, G., Riley, S., van Elsland, S., Volz, E.,
  Wang, H., Wang, Y., Xi, X., Donnelly, C.~A., Ghani, A.~C., and Ferguson,
  N.~M. (2020).
\newblock Estimates of the severity of coronavirus disease 2019: a model-based
  analysis.
\newblock {\em Lancet Infectious Diseases}, 20(6):669--677.

\bibitem[Viana et~al., 2022]{viana2022rapid}
Viana, R., Moyo, S., Amoako, D.~G., Tegally, H., Scheepers, C., Althaus, C.~L.,
  Anyaneji, U.~J., Bester, P.~A., Boni, M.~F., Chand, M., et~al. (2022).
\newblock Rapid epidemic expansion of the {SARS-CoV-2} {O}micron variant in
  southern {A}frica.
\newblock {\em Nature}, 603(7902):679--686.

\bibitem[{Waitangi Tribunal}, 2021]{waitangi2021haumaru}
{Waitangi Tribunal} (2021).
\newblock Haumaru: The {COVID}-19 priority report, {WAI2575}.
\newblock waitangitribunal.govt.nz/assets/Covid-Priority-W.pdf.

\bibitem[Ward et~al., 2022]{ward2022risk}
Ward, I.~L., Bermingham, C., Ayoubkhani, D., Gethings, O.~J., Pouwels, K.~B.,
  Yates, T., Khunti, K., Hippisley-Cox, J., Banerjee, A., Walker, A.~S., et~al.
  (2022).
\newblock Risk of {C}ovid-19 related deaths for {SARS-CoV-2} {O}micron
  ({B.1.1.529}) compared with {D}elta ({B.1.617.2}): retrospective cohort
  study.
\newblock {\em British Medical Journal}, 378:e070695.

\bibitem[Zarebski et~al., 2017]{zarebski2017model}
Zarebski, A.~E., Dawson, P., McCaw, J.~M., and Moss, R. (2017).
\newblock Model selection for seasonal influenza forecasting.
\newblock {\em Infectious Disease Modelling}, 2:56--70.

\end{thebibliography}
\end{document}